\documentstyle[aps,twocolumn,epsf,citesort]{revtex}

\begin{document}
\title
{Transitions between Inherent Structures in  Water}

\author{Nicolas Giovambattista$^1$, Francis W. Starr$^2$, 
Francesco Sciortino$^3$,\\ 
Sergey V. Buldyrev$^1$,
H. Eugene Stanley$^1$}

\address{$^1$Center for Polymer Studies and Department of Physics,
  Boston University,\\Boston, Massachusetts 02215.}
\address{$^2$Polymers Division and Center for Theoretical and
  Computational Materials Science, National Institute of Standards and
  Technology, Gaithersburg, MD, 20899 USA} 
\address{$^3$
Dipartimento di Fisica, Istituto Nazionale per la Fisica della
Materia,\\
and I.N.F.M. Center for Statistical Mechanics and Complexity,\\
Universit\`{a} di Roma La Sapienza, P.le A. Moro 2, I-00185 Roma, ITALY}

\date{EH8157. Time scale corrected in Figs 1,2,8: 31 Dec 2001}

\maketitle

\begin{abstract}
The energy landscape approach has been useful to
help understand the dynamic properties of supercooled liquids 
and the connection between these properties and 
thermodynamics.  The analysis in numerical models of the inherent
 structure (IS) trajectories ---
the set of local minima visited by the liquid ---
offers the possibility of filtering out the vibrational component of the
motion of the system on the potential energy surface 
and thereby resolving the slow structural component more efficiently.
Here we report an analysis of an IS trajectory for a
widely-studied water model, focusing on the changes in hydrogen bond
connectivity that give rise to many IS separated by
relatively small energy barriers.  We find that while the system
\emph{travels} through these IS, the structure of
the bond network continuously modifies, exchanging linear bonds for
bifurcated bonds and usually reversing the exchange to return to nearly
the same initial configuration. For the 216 molecule system we investigate,
 the time scale of these transitions is as small as the simulation
time scale ($\approx 1$~fs).  Hence for water, the transitions between
each of these IS is relatively small and eventual
relaxation of the system occurs only by many of these
transitions.  We find that during IS changes, the molecules
with the greatest displacements move in small ``clusters'' of 1-10
molecules with displacements of $\approx 0.02-0.2$~nm, not unlike
simpler liquids. However, for water these clusters appear to be somewhat more
branched than the linear ``string-like'' clusters formed in a supercooled Lennard-Jones
 system found by Glotzer and her collaborators.

\end{abstract}


\section{Introduction}

It is believed that the properties of liquids can be
understood as motion of the system in a high-dimensional complex
potential energy surface (PES)
\cite{Goldstein,stillweber,still95,Angell95,debenedetti,newnat}. As a
liquid is cooled toward the glassy state, the system is increasingly
found near local potential energy minima, called  inherent
 structure (IS)
 configurations \cite{stillweber}.  As temperature decreases,
 the description of the dynamics in terms of motion on the
PES becomes increasingly appropriate. In this description, in
 the glassy state, the system is localized in one of the potential energy
basins \cite{schroder,heuer,angelani,francescokob}.

While such a picture of liquid dynamics is difficult to verify
experimentally, computer simulation offers an excellent opportunity to
explore these ideas.  For a pre-defined liquid potential, a liquid
trajectory can be generated via molecular dynamics simulation and the
 local potential energy minima can
be evaluated by an energy minimization method \cite{stillweber}.
 With this procedure, the
motion in phase space is converted into a minimum-to-minimum trajectory,
or {\it IS trajectory}. A general picture of the system moving
among a set of basins surrounding the multitude of local minima has 
evolved.  More specifically, simulations have shown that both the
depth of the minima sampled by the system, as well as the number of
these minima, decrease on cooling \cite{debenedetti,francescokob,nature}.
Simulations have also shown that below 
a crossover temperature $T_\times$, only rare fluctuations
bring the system to a saddle point and hence activated processes become
important for relaxation of the liquid
\cite{schroder,angelani,wales,keyes,lanave}.
  For the system we study,  $T_\times$ coincides numerically 
\cite{angelani,lanave,st,claudio,lanavesilica}
with the critical
temperature identified by mode coupling theory (MCT), which has
 been widely used to understand the dynamics of liquids on ``weak''
supercooling (the $T$ range where characteristic relaxation times
approach $\approx 10^{-8}$~s) \cite{mct}.

The description of the real motion of the system as an IS trajectory
becomes a powerful way of separating the vibrational contribution,
responsible for the thermal broadening of instantaneous measurements
from the slow structural component\cite{tanakaohmine}. Such
an approach becomes even more powerful below $T_\times$, since most of the instantaneous
configurations are far from saddles, making correlation functions
calculated from the IS trajectory fully account for the
$\alpha$-relaxation dynamics\cite{schroder}.

Here we study the IS trajectory below $T_\times$ for the extended
simple point charge (SPC/E) potential\cite{spce}, a well-studied model for
water.  The dynamics of the SPC/E model have been
shown to be consistent with the predictions of
MCT~\cite{francescos,francislong}.  Additionally, the PES of SPC/E
 has been studied in detail above $T_\times$ and a thermodynamic
description of the supercooled states based on the PES has been
presented\cite{nature,francissastry}. Here we focus on the geometrical
properties of the motion, once the vibrational component is subtracted.
 The possibility of performing such a study below $T_\times$, with a
very fine time coarse graining, allows us to examine the
structural changes that accompany the basin
transitions and to describe an elementary step of the diffusive process in
terms of hydrogen bond network rearrangement.

This work is organized as follows. In Sec.~\ref{Simulation} we provide the
simulation details. In Sec.~\ref{IS} we analyze the displacement 
of the molecules in these IS-transitions and in Sec.~\ref{HB} we study the
corresponding changes in the hydrogen-bond (HB) network. 
Finally, in Sec.~\ref{conclusions} we present a brief summary.

\section{Simulation}
\label{Simulation}

Our results are based on molecular dynamics simulations of the SPC/E model
 \cite{spce} of water for 216 molecules, at fixed density $\rho =
 1$~g/cm$^3$~.  The numerical procedure is described in Ref.\cite{francislong}.
The integration time step $\delta t$ is $2$ fs.
 The mode coupling temperature for this density is
$T_{MCT} \approx 194K$ \cite{francislong}. We analyze trajectories at
 $T=180$~K, so that the system is in the deep supercooled liquid
state. At this temperature, the diffusion coefficient is more than four
orders of magnitude smaller than its value at $T=300$~K and
only a few molecules move significantly (with displacements larger than $0.025$~nm)
 at each simulation time step\cite{extra0}. Our system is started
 from equilibrated configurations at $190$~K, which relax
 for nearly 920~ns at $180$~K before we record and analyze the
 trajectory. At such a low temperature, a slow aging in the trajectory could be
 present, however, the aging should not affect the qualitative
picture we present.

We have generated one trajectory of $30$~ns, sampling configurations
at each $1$~ps. For each configuration we find the
corresponding IS using conjugate gradient minimization.  In this way, we
obtain $30,000$ configurations with the corresponding IS.
Since we could miss some IS transitions with $1$~ps sampling, we
 also ran four independent $20$~ps simulations sampling the IS at $4$~fs. In this way,
 we obtain another 20,000 configurations with the corresponding IS.

\section{Inherent Structure Trajectories}
\label{IS}
Figure~\ref{PE} shows an example of the potential energy of the IS,
$E_{IS}(t)$, and the mean square displacement of the oxygen atoms
$\langle r^2(t) \rangle$ starting from a single arbitrary starting time.
 At $T=180$~K, the slowest collective relaxation time $\tau_{\alpha}
> 200$~ns \cite{francislong}.  
The IS trajectory in Fig.~\ref{PE} has a mesh of $1$~ps and covers a total
 time of $30$~ns. In this time interval, $\langle
r^2(t)\rangle$ is about $1\AA^2$, i.e. much less than the corresponding
value of the average nearest
neighbor distance of $2.8 \AA$. Figure~\ref{Peak} shows an enlargement of
the IS trajectory using a much smaller time mesh ($4$~fs,
two times the simulation time step).  Fig.~\ref{Peak} shows that changes occur via
discrete transitions, with an average duration of $\approx 0.2$~ps.
The transitions are characterized by an energy change of $\approx
10-20$~kJ/mol and an oxygen atom square displacement of the order of
$0.01~\AA^2$; they appear to constitute the elementary step underlying the
diffusional process in the system.

We note that it is impossible to identify the transition unless the
quenching time step is of the order of the simulation time step.  This
is in distinct contrast to a Lennard-Jones liquid, where such small
continuous changes are not present~\cite{schroder}. The difference, we
will show, is attributable to the hydrogen bonds. This time scale
 for IS transitions is in accord with other work
on the TIPS2 water model, at $T=298$~K \cite{tanakaohmine}.

In Fig.~\ref{Peak}, we see the sharp changes in $E_{IS}(t)$ coincide with
the sharp changes in $\langle r^2(t) \rangle$.  This confirms that the system is repeatedly
visiting specific configurations, since $\langle r^2(t) \rangle$ and $E_{IS}(t)$ take on
discrete values.  The results shown in Fig.~\ref{Peak} imply that
 the system often returns to the original basin because of both
the difference in energy and the displacement approach zero at the end of the
time interval.

To aid in understanding the distribution of the displacements
during the IS changes (such as those between the two IS labeled A and B
in Fig.~\ref{Peak}(b)), Fig.~\ref{dri} shows the
displacements $u$ of all 216 individual molecules from IS-A to IS-B. We see that
there is a relatively small set of molecules with a large displacement.
A  snapshot of the eight molecules with the largest displacement is shown in
Fig.~\ref{snapshot1}. Interestingly, we find that this set of molecules forms a cluster
of bonded molecules. Indeed, for all cases studied, 
we found that the set of molecules which displace most form a cluster
of bonded molecules. 
The observed 
clustering phenomenon characterizes the IS transitions in water
and can be interpreted as the analog of the 
string-like motion observed in simple atomistic liquids~\cite{schroder} 
and connected
to the presence of dynamical heterogeneities\cite{dyneth}.
Similar results were found by Ohmine {\it et} al.
using the TIP4P and TIPS2 models for water~\cite{ohmineandref}.

To characterize the distribution of individual molecular displacements 
between different IS more carefully, we show 
the distribution of displacements $u$ of the oxygen atoms
$P(u,\delta t=4~\mbox{fs})$ in Fig.~\ref{dr180}. The time difference $\delta t$
is two times the integration time step\cite{note}. 
We collected data from
 all transitions in four independent $10$~ps trajectories for the results 
reported in Fig.~\ref{dr180}. 
Note that $P(u)$ was  previously studied by
Schr{\o}der {\it et al.} for a binary Lennard-Jones (LJ) mixture ~\cite{schroder}.

We find that the distribution of displacements is monotonic and does not allow us  
to detect a characteristic length
which could help in distinguishing between diffusive and non-diffusive
basin changes.  
$P(u, t)$ for $t=\delta t$ shows an apparent power-law with a negative slope of about
$-2.8\pm0.2$, followed by an exponential tail. 
To highlight the exponential tail, we present the data in a linear-log plot
Fig.~\ref{dr180} (upper panel), showing that 
the tail in $P(u,\delta t=4~\mbox{fs})$ is mostly due to these highly ``mobile molecules''.
The  characteristic length of the exponential tail is 
$0.2 \AA$, about 
15 times smaller than the nearest neighbor oxygen distance of  $2.8 \AA$ and 
in agreement with the finding of Schr{\o}der et al. for the LJ case,
where the characteristic length  was about $1/8$ of the 
nearest neighbor distance.

Schr{\o}der {\it et al.} interpreted the
power-law contribution to $P(u, \delta t)$ as a response of the
system to the diffusing molecules. In elasticity theory\cite{dyre},
a local displacement at the origin produces a distribution of
displacements which scales with the distance $R$ from the origin 
as $R^{-2}$. Hence the distribution of displacements $u$
 scales as $u^{-2.5}$\cite{dyre}. In contrast with the LJ case, 
the exponent we find is slightly larger than $2.5$, a discrepancy 
which may be related to the application of a 
continuum theory at a length scale where molecular details are
still relevant. In the spirit of elasticity theory,
the deviation of the distribution at small $u$ values from the power
 law arises from the
cutoff introduced by the finite size of the simulation.
Indeed, small $u$ values are produced at large $R$
 and hence are missing in a finite simulation box.

Although a subset of ``highly mobile'' molecules is identifiable using
a pre-defined threshold value in a single basin change,
there is no unambiguous general criterion for identifying 
the molecules responsible for a single basin transition.

  An important open question is the relation between
the displacement distribution functions for $t=\delta t$ (Fig.~\ref{dr180})
 and the same distribution evaluated for an arbitrary time interval $t$. 
A full knowledge of such relation may shed light on the elementary
stochastic process which better describes the dynamics of the deep
supercooled state.  This calculation requires a significant
computational investment due to the very large number of IS
configurations which need to be evaluated. Despite such difficulties,
it is a promising research line for the near future.

\section{HB network restructuring}
\label{HB}
Liquid water is an interesting system for studying the physical processes
that accompany basin transitions. The
unusual dynamic and thermodynamic properties of water are believed to be
connected to the microscopic behavior of hydrogen bonding.  Many
experiments suggest that this tetrahedral HB network has defects, such
as an extra (fifth) molecule in the first coordination
shell\cite{texeira,gngw}.  Indeed, such 5-coordinated molecules have
been directly identified in simulations and the defects were found to
be a catalyst for motion in the system\cite{geiger}, making an obvious
possible connection between network defects promoting diffusion and the
basin transitions that give rise to diffusive motion of that
system.

To obtain a physical picture of the IS transitions for water and to
hopefully better understand the source of these transitions, we focus
on the small changes in the HB network along a minimum to minimum trajectory.
 Analysis of the HB network based on IS configurations has shown that local PES
minima are characterized by both linear bonds (LB) and bifurcated bonds (BB)
 \cite{geiger}
 whose fraction is both temperature and density dependent.  The HB network
tends toward a perfect random tetrahedral network on cooling, or on
lowering the pressure\cite{francissastry}. Here we emphasize the
changes in the network associated with IS-transitions.

To quantify the bonding changes, it is necessary to employ an
arbitrary bond definition.  Previous works have indicated that several
definitions provide physically reasonable results.  Here we use the
definition that two molecules are bonded if their oxygen-oxygen distance
is less than 3.5\AA~and their mutual potential energy is negative
\cite{geiger}. Using this definition, we obtain
the equilibrium distribution of the potential energy for the LB and BB,
 shown in Fig.~\ref{Vij}.

 These results agree with previous findings, based on different
models\cite{st2,spc,tip4p}.  The distribution of BB energies is bimodal
with peaks at roughly $-6$~kJ/mol or $-22.5$~kJ/mol, while the LB energy
distribution is unimodal with a peak at roughly $-24$~kJ/mol.
Therefore, the energy associated with a change in the HB network due to
losing {\it one} LB and creating {\it two} BB ranges roughly from
$-21$~kJ/mol to $12$~kJ/mol, depending on which of the two possible BB are
created. The relative intensities of the peaks of the
BB energy distribution suggest that such a mechanism would more likely
lead to an increase in the overall energy. The comparison of the HB-energy changes
 with those found in Fig.~\ref{PE} suggest that interchanges between LB and
 BB can explain the
existence of a multitude of IS separated by very small energy barriers.
More specifically, we hypothetize that changes increasing $E_{IS}$
 found in Fig.~\ref{PE} are
due to processes LB$\rightarrow$BB, while the changes decreasing $E_{IS}$
 are due to the processes BB$\rightarrow$LB.
To confirm or reject this hypothesis, we study the distributions of LB
and BB for IS just before and just after positive and negative ``jumps'' in
the potential energy (Fig.~\ref{PE}b) with energy larger
 than 9~kJ/mol \cite{comment}. These IS are schematically shown in
Fig.~\ref{LB-BB}(a).  The corresponding distributions for LB
and BB are shown in Fig.~\ref{LB-BB}(b) and~\ref{LB-BB}(c). We see that
during PE increases, the average number of LB decreases while the number
of BB increases; the opposite situation occurs when the potential energy
decreases.  The distribution for BB has peaks at even numbers of BB,
which we expect since for each LB lost, two BB appear, also implying
that the distribution should be zero for odd numbers of BB \cite{noteBB}.
 The anti-correlation of the BB and LB changes is a
strong indicator that a mechanism whereby the system accesses
higher energy states is via LB$\rightarrow$BB transitions.

To reinforce the hypothesis that the basin change is associated with a
restructuring of the local connectivity, Fig.~\ref{CN} shows the number
of molecules with a coordination number equal to 3, 4 or 5 as a function of
time for a characteristic time interval and contrasts this data with
the time dependence of $\langle r^2(t) \rangle$.  A clear anti-correlation is observed
between the time dependence of the number of 3 and 5-coordinated molecules compared to
the time dependence of the 4-coordinated molecules, supporting the proposed 
interchange mechanism. The fact that
increases in the fraction of BB coincides with changes in 
 $\langle r^2(t) \rangle$ supports the expectation that motion is
a result of network imperfections.

\section{Summary}
\label{conclusions}
We have presented a detailed analysis of the motion on the PES of
 a 216 water molecule system interacting via the SPC/E potential in a
deeply supercooled state, below the MCT temperature.  At this
temperature, the system populates basins of the local minima.
 At these conditions, the analysis
of the IS trajectories provides a very clean description of 
the slow alpha relaxation process filtering nearly all vibrational motions.

We have shown that an inherent structure transition is observed about
every 0.2 ps. It is the collection of these numerous small
transitions that gives rise to the structural relaxation of the system.
These fast transitions are characterized by a broad
distribution of individual molecule displacements, without a clear
characteristic length. 
Future work must address
the issue of the prediction of $P(u,t)$ from the knowledge of
$P(u,\delta t)$.

We perform an analysis of the geometry of the individual event and find 
 that the most mobile molecules are clustered. The
analysis of the changes in HB connectivity associated with IS changes
reveals that these transitions are associated with the breaking and
reformation of HB. This result is in accord with work of 
Ohmine and colleagues \cite{tanakaohmine,ohmineandref} for TIP4P.
 We have shown that the transitions associated
with an increase in the energy correspond to the breaking
of linear bonds and to the simultaneous formation of 
bifurcated bonds. Similarly, the transitions associated
with a decrease in the energy correspond to the breaking
of bifurcated bonds and to the simultaneous formation of 
linear bonds.

 This result supports the hypothesis that the
linear to bifurcated transition can be considered as an elementary step
in the rearrangement of the HB network.

\section{Acknowledgments}
We thank T.B. Schr{\o}der, S. Sastry and S.C. Glotzer 
for interesting discussions. N.G. wants to thanks I. Ohmine
for fruitful discussions and suggestions.
F.S. acknowledges support from INFM-PRA-HOP and Initiaziva
Calcolo Parallelo and MURST COFIN2000.
This work was supported by the NSF Chemistry Program.

\begin{figure}[htb]
\narrowtext \centerline{
\hbox {
 \vspace*{0.5cm}  
  \epsfxsize=10cm
  \epsfbox{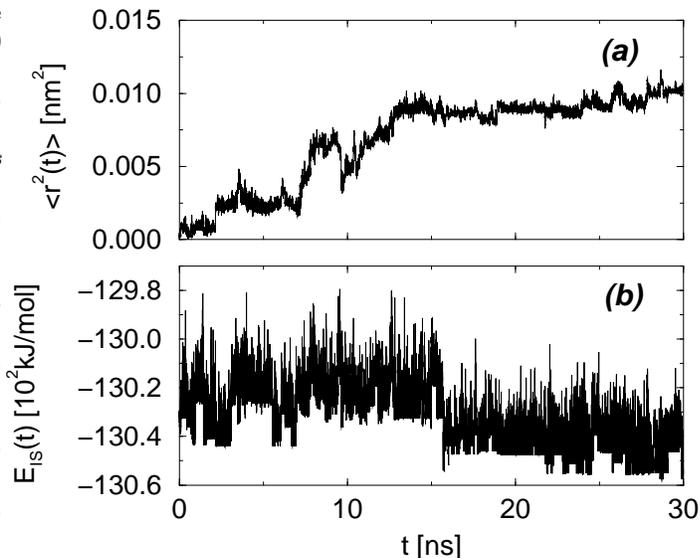}
  }
   }

\vspace*{0.5cm}

\caption{(a) Mean square displacement and (b) IS energy 
for the inherent structures  as a function of time for
the studied 216-molecule system. The
  time interval between adjacent IS in both figures is $1$~ps.
 While it is possible to track
  IS transitions from the potential energy, it is not the case for the 
  mean square displacement. Note the amplitude of
  the peaks of the potential energy is $\approx 10-20$~kJ/mol, the
  same order of magnitude of a HB energy.}

\label{PE}
\end{figure}   

\vspace*{1.0cm}

\begin{figure}[htb]
\narrowtext
\centerline{
\hbox {
  \vspace*{0.5cm}  
  \epsfxsize=9.5cm
  \epsfbox{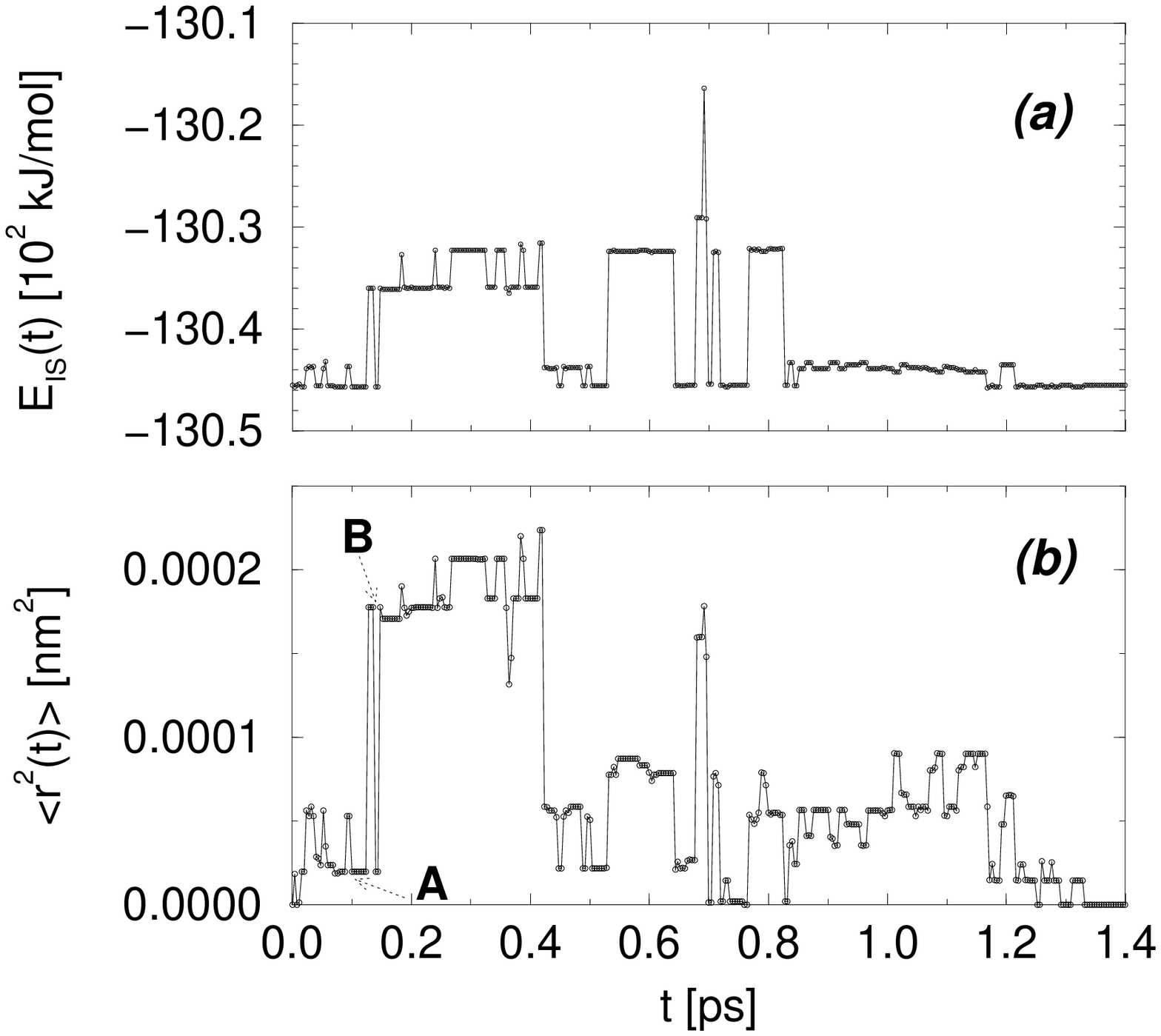}
  \hspace*{0.3cm}
  }
   }

\vspace*{0.5cm}

\caption{(a) IS energy and (b) mean square displacement
 for the IS obtained using a sampling interval of $4$~fs, two times
 the simulation time step. The correlation
  between $E_{IS}$ and $\langle r^2(t) \rangle$ is evident. Also, we
 see that it is necessary to sample IS with a mesh of the order 
of the simulation time step to detect all the IS
  visited by the system.}
\label{Peak}
\end{figure}

\vspace*{1.0cm}

\begin{figure}[htb]
\narrowtext \centerline{
\hbox {
  \vspace*{0.5cm}  
  \epsfxsize=9.5cm
  \epsfbox{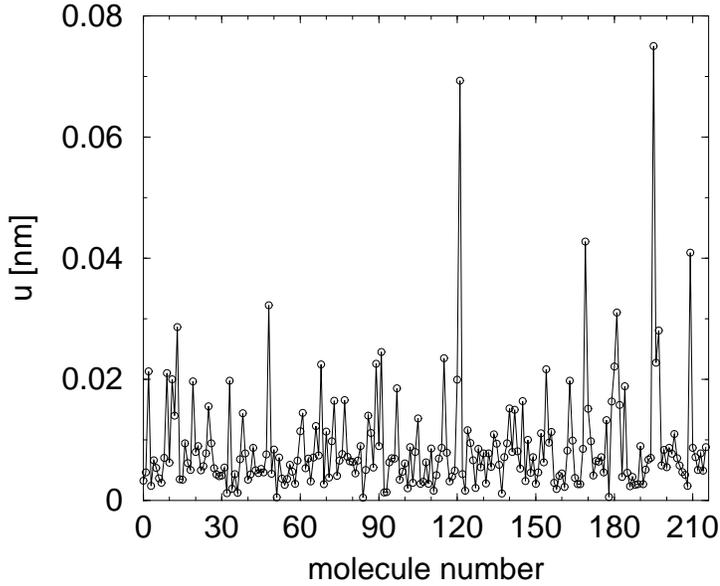}
  \hspace*{0.3cm}
  }
   }

\vspace*{0.5cm}

\caption{Displacement of each molecule in the transition from IS-A to IS-B
  shown in Fig.~\ref{Peak}(b).}
\label{dri}
\end{figure}   

\vspace*{1.0cm}

\begin{figure}[htb]
\narrowtext \centerline{
\hbox {
  \vspace*{0.5cm}  
  \epsfxsize=8cm
  \epsfbox{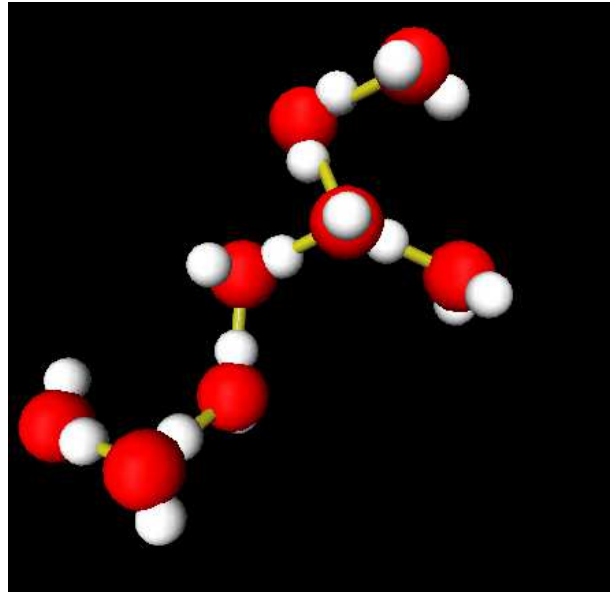}
  \hspace*{0.3cm}
  }
   }

\vspace*{0.5cm}

\caption{Snapshot of the system in the IS labeled A in Fig.~\ref{Peak}(b).
 Only the eight molecules with displacement larger than $0.025$~nm
  are shown here. Hydrogen-bonded molecules are connected by tubes. Note that
  all 8 molecules are nearby and form a cluster, which unlike the LJ case,
  are bonded and less string-like.}
\label{snapshot1}
\end{figure}

\vspace*{1.0cm}

\begin{figure}[htb]
\narrowtext 

\centerline{
\hbox {
  \vspace*{0.5cm}  
  \epsfxsize=9.5cm
  \epsfbox{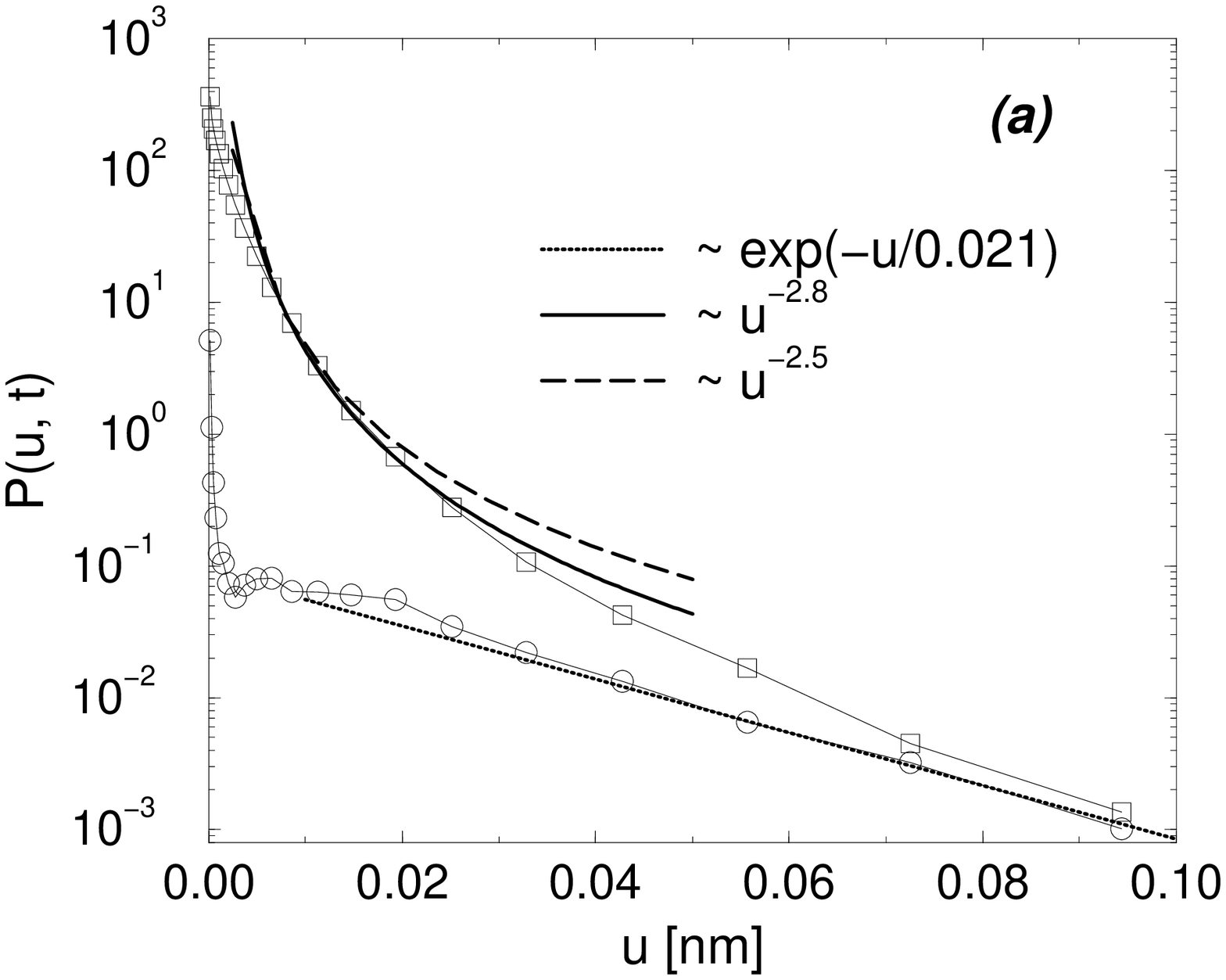}
  \hspace*{0.3cm}
  }
   }

\centerline{
\hbox {
  \vspace*{0.5cm}  
  \epsfxsize=9.5cm
  \epsfbox{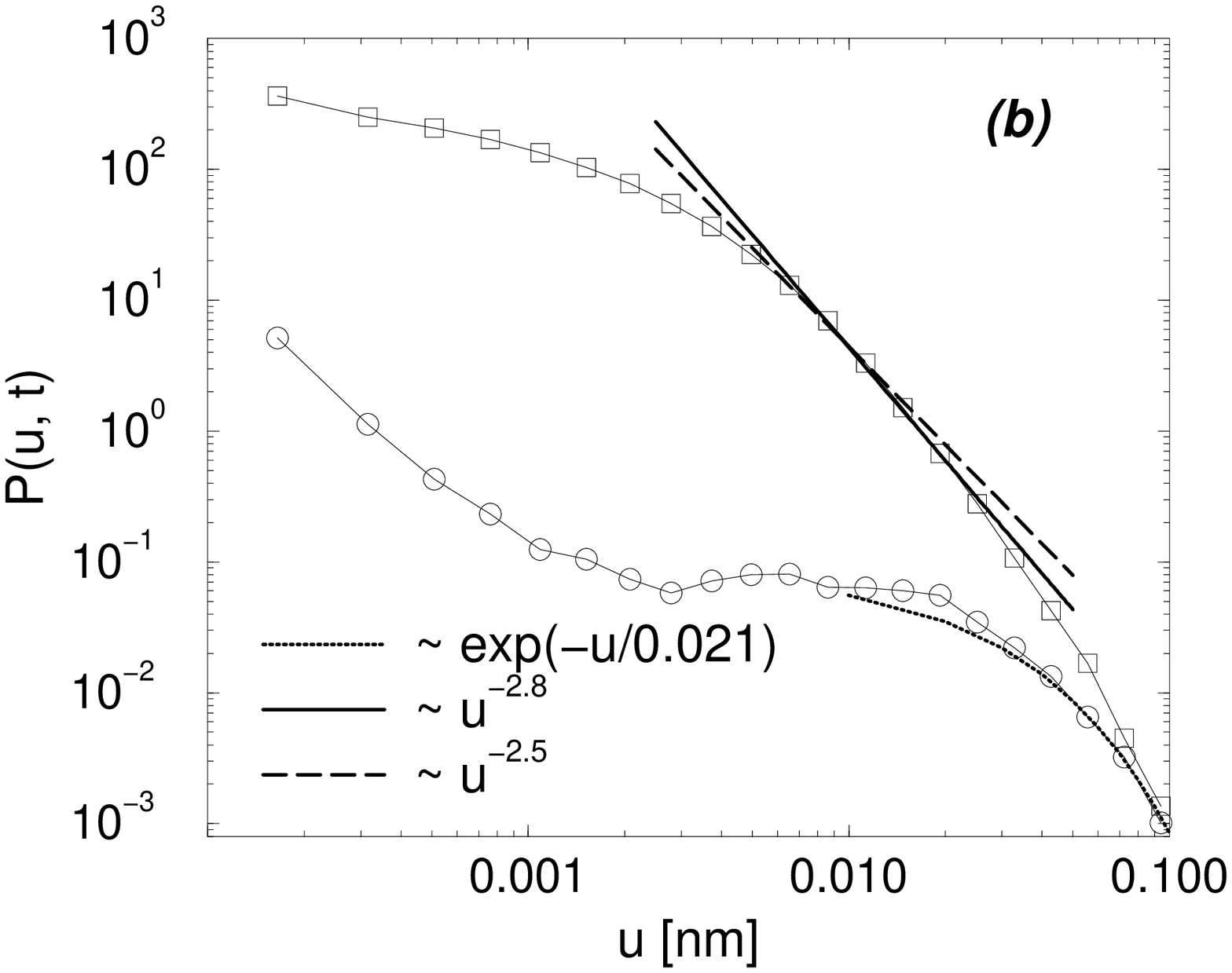}
  \hspace*{0.3cm}
  }
   }

\vspace*{0.5cm}

\caption{Distribution of displacements $u$ of the oxygen atoms 
 between IS changes, $P(u,t)$, evaluated at $t=4$~fs. We show the distribution of all the
 displacements of the molecules (squares) and the distribution for the
 largest displacement in an IS-transition (circles). For comparison, we
 divided the largest displacement distribution by the number of molecules.
 We present both (a) log-log and (b) linear-log graphs to show
 the power law behavior for $u \approx 0.01$~nm and the exponential tail
 of the distributions. We also show the prediction of elasticity theory
 $P(u) \sim u^{-2.5}$.}
\label{dr180}
\end{figure}

\vspace*{1.0cm}

\begin{figure}[htb]
\narrowtext
\centerline{
\hbox { 
  \vspace*{0.5cm} \epsfxsize=10.0cm \hspace*{0.25cm} 
  \epsfbox{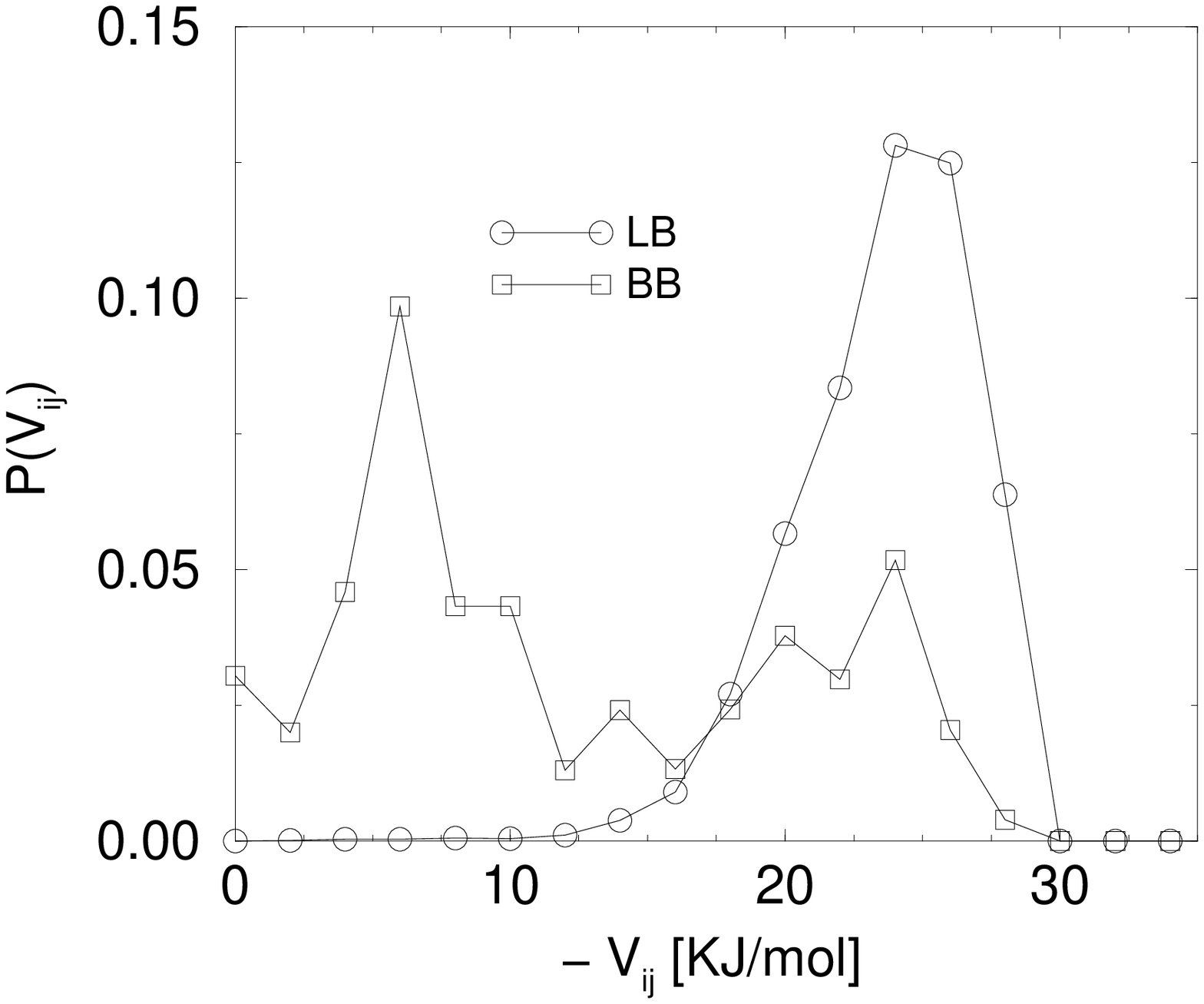} }
  }

\vspace*{0.5cm}

\caption{Distribution function for the pair interaction energy
  $V_{ij}$ for the LB and BB in the IS. Note
  the bimodal distribution for BB and unimodal distribution for LB.}
\label{Vij}
\end{figure}

\vspace*{1.0cm}

\begin{figure}[htb]
\narrowtext

\centerline{
\hbox {
  \vspace*{0.5cm}
  \epsfxsize=8.5cm
  \hspace*{1.0cm}
  \epsfbox{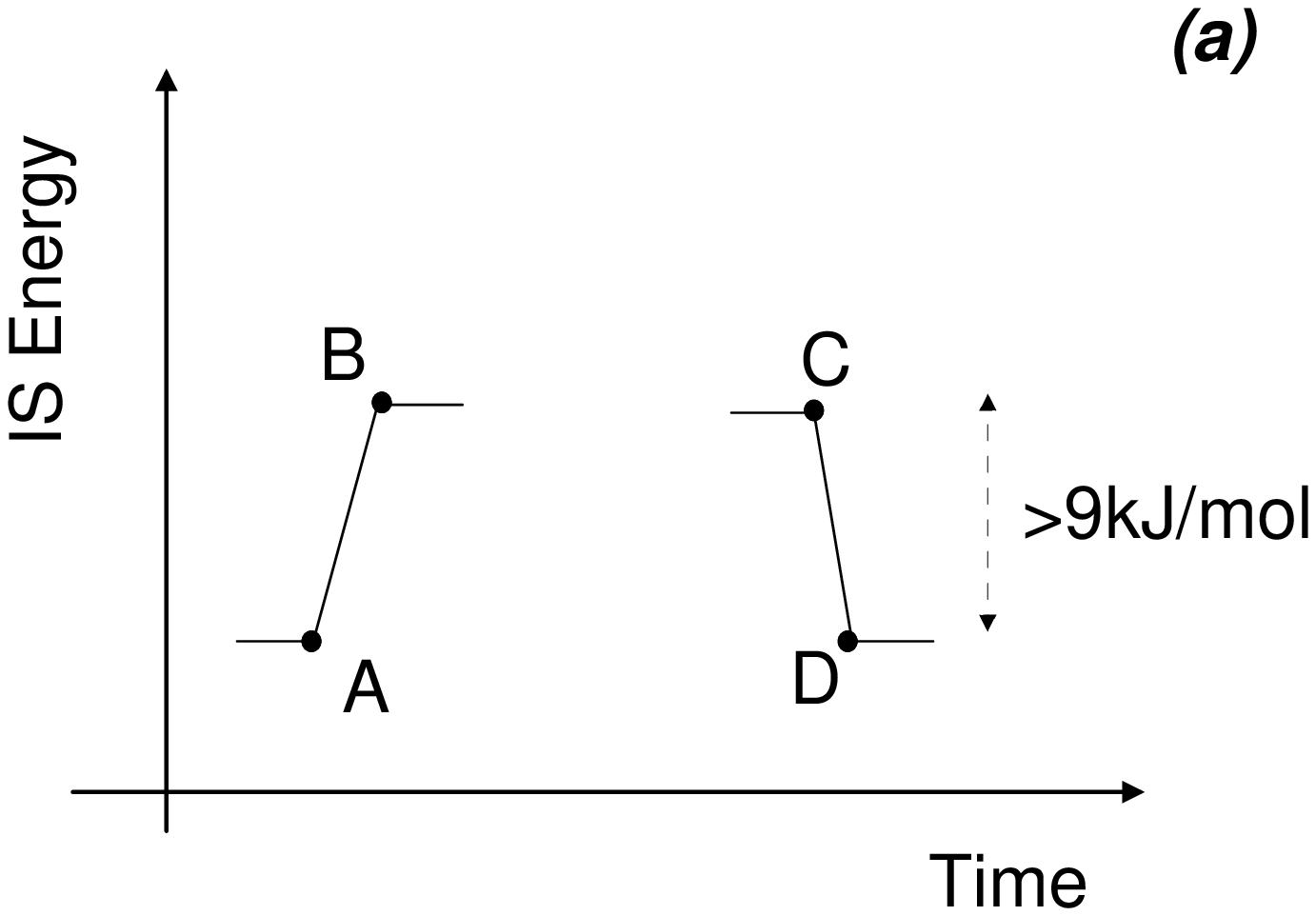} 
        }
         }
\vspace*{0.5cm}
\centerline{
\hbox {
  \vspace*{0.5cm}
  \epsfxsize=9cm
  \epsfbox{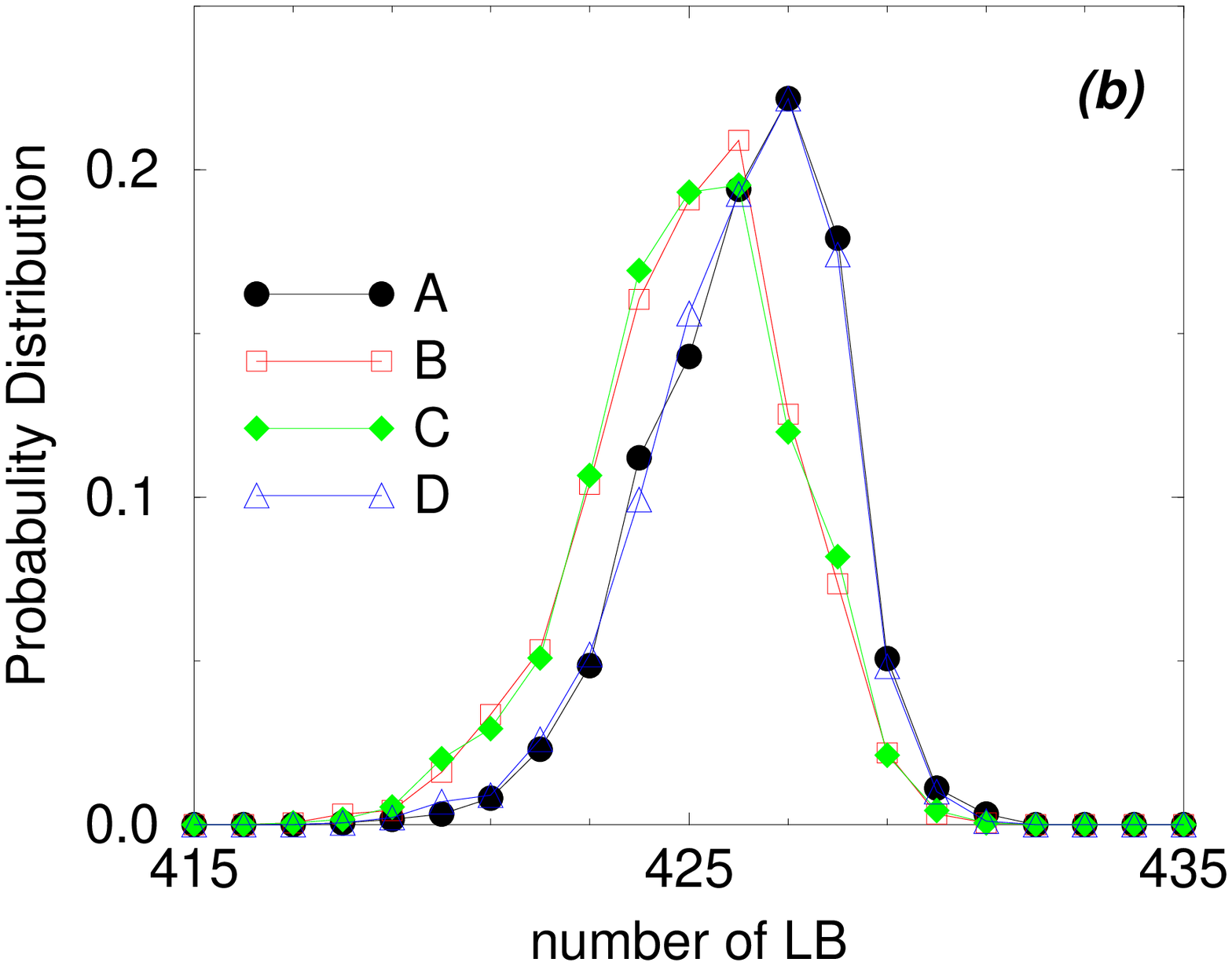} 
        }
         }
\vspace*{0.01cm}
\centerline{
\hbox {
  \vspace*{0.5cm}  
  \epsfxsize=9cm
  \epsfbox{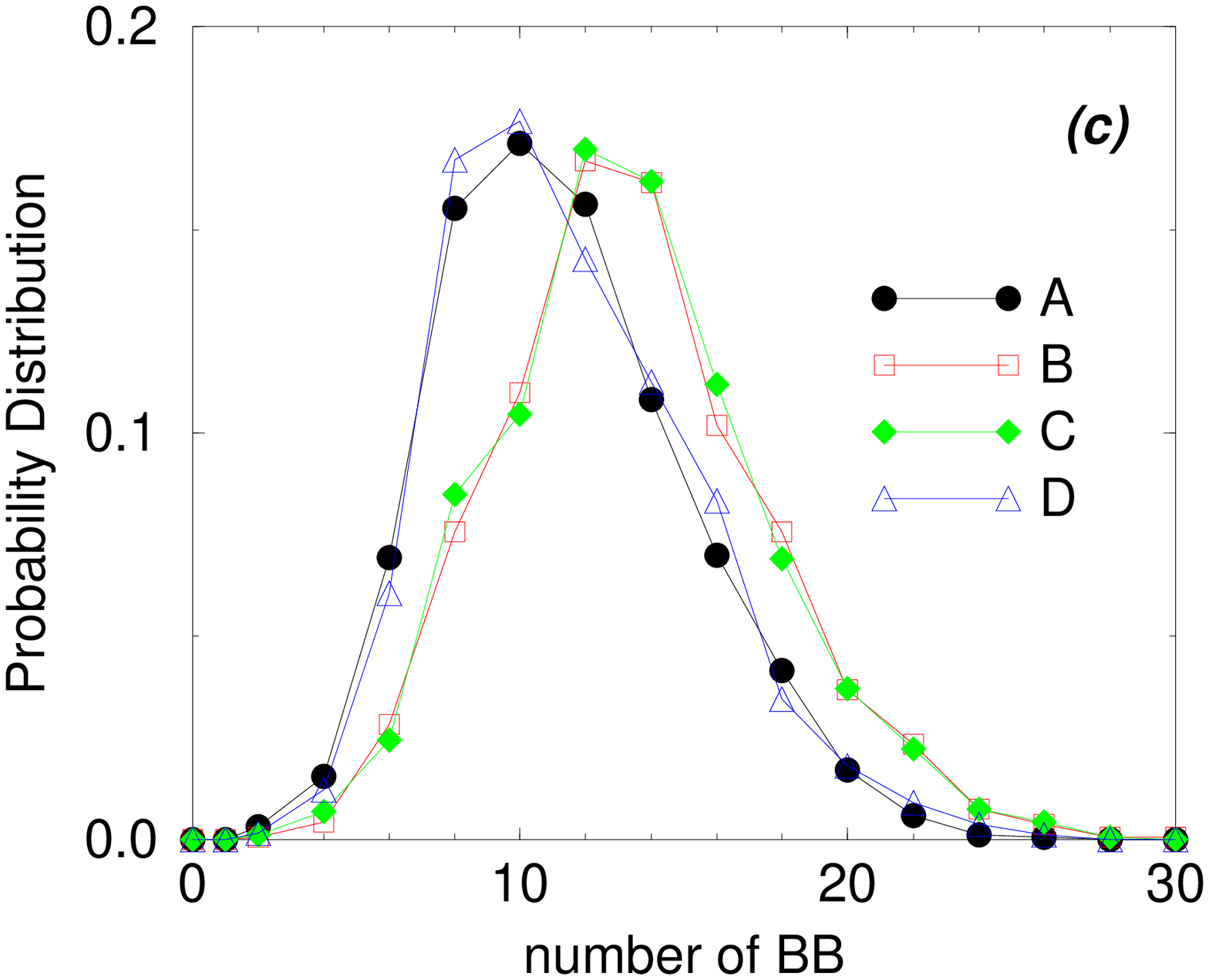}
  }
   }
\vspace*{0.5cm}

\caption{Panel (a) shows schematically the separation of IS used to meassure
 the probability distributions for (b) LB and (c) BB.
 Circles correspond to the distribution calculated over IS
just before an increase in potential energy larger than 9 kJ/mol, i.e. when
 the system is in an IS A shown in (a). Squares correspond to the
 distribution calculated over IS just after an increase in potential energy
 larger than 9 kJ/mol, i.e. when
 the system is in an IS B shown in (a).
Similarly, diamonds correspond to the distribution calculated over IS
just before a decrease in potential energy larger than 9 kJ/mol, when
 the system is in an
IS C shown in (a). Triangles correspond to the
 distribution calculated over IS
just after a decrease in potential energy larger than 9 kJ/mol, when
 the system is in an IS D shown in (a).}   

\label{LB-BB}
\end{figure}   

\vspace*{1.0cm}

\begin{figure}[htb]
\narrowtext \centerline{
\hbox {
  \vspace*{0.5cm}  
  \epsfxsize=9cm
  \epsfbox{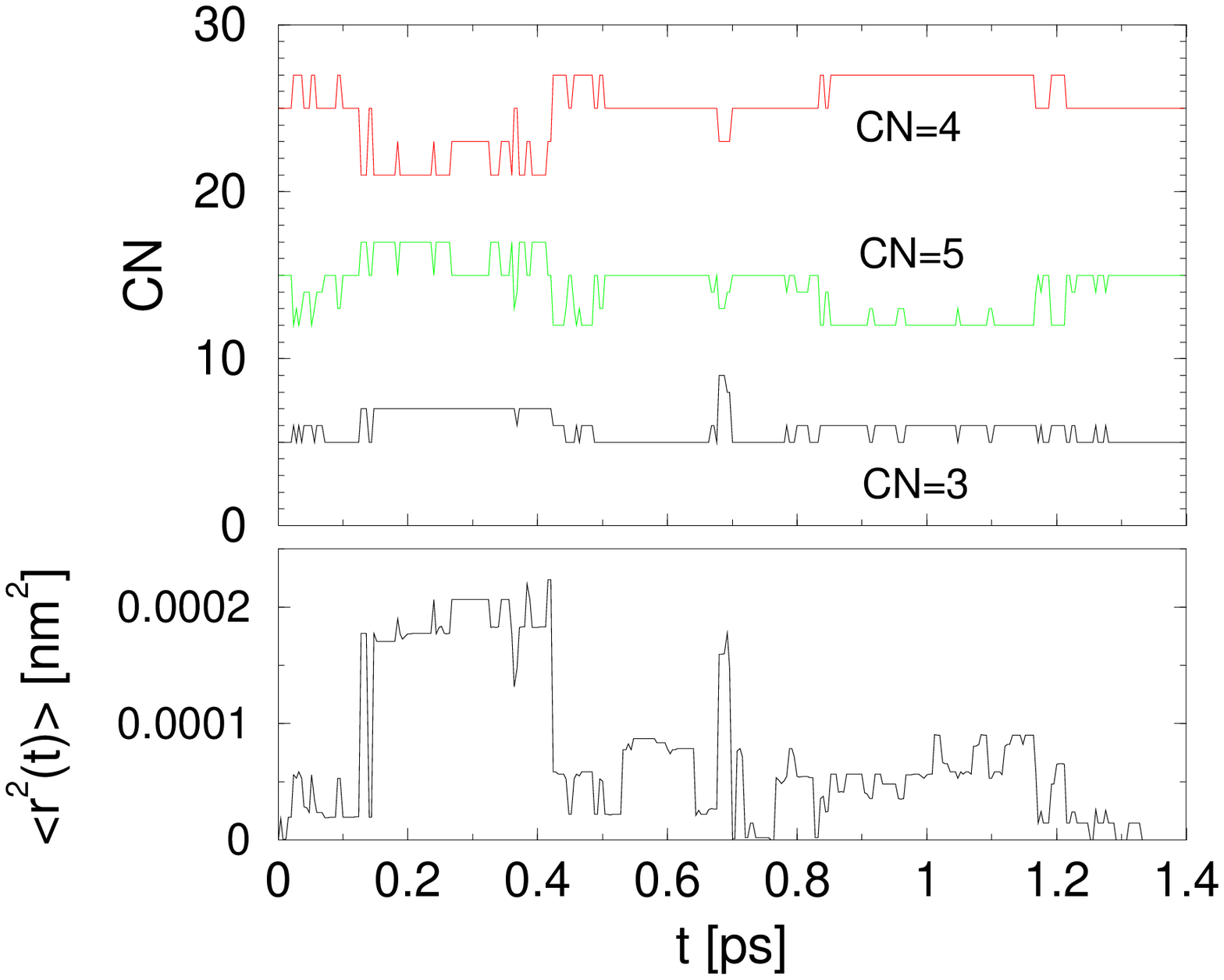}
  \hspace*{0.3cm}
  }
   }

\vspace*{0.5cm}

\caption{Number of molecules with coordination number CN equals 3,4 and
  5 versus time, for the IS corresponding to Fig.~\ref{Peak}. The plot for
  $CN=4$ is shifted down by 400 units for better comparison. Also shown is 
$\langle r^2(t) \rangle$. We see how the tetrahedral network acquires both
types of defects (CN equals to 3 and 5) while the system explores different IS.}
\label{CN}
\end{figure}


\begin{references}
\bibitem{extra0}
The Adam and Gibbs hypothesis \cite{extra1} states that flow in a supercooled fluid
 involves cooperative motion of molecules.
For recent experimental work on cooperative motions see Ref.\cite{extra2}.
\bibitem{note} If during the time interval  $\delta t$ 
the system does not change IS, the contribution to $P(u,\delta t)$ 
appears only for $u=0$ and hence does not modify the shape of
the distribution.
\bibitem{comment}
We find from Fig.~\ref{PE} that most of the ``jumps'' in potential energy 
correspond to changes in $|E_{IS}|$ larger than 9kJ/mol. We found
similar results for a lower cutoff in energy of 1kJ/mol.
\bibitem{noteBB}
We find that the distribution is not exactly zero for odd numbers of
BB, which is a consequence of the bond definition.
Using the definition that two molecules are bonded if their
oxygen-oxygen distance is less than $0.35$~nm and their interaction energy is
less than a fixed value $E_{HB}=0$, we find that there exist few cases in which
one hydrogen atom is ``shared'' by 3 oxygens (instead of 2 as in the case
of a BB).  We try different values for $E_{HB}$ and we find that
these ``tri-furcated'' bonds (TB) disappear if $E_{HB}$ is reduced to
-10kJ/mol. These TB can also contribute to the
presence of an even number of BB appearing in the distribution. The
ratio of BB:TB for $E_{HB}=0$, $-5$~kJ/mol is 20:1 and
24:1.  We also check that the distribution for BB
at $E_{HB}=-10$ kJ/mol is zero for odd numbers of molecules.
\end{references}
\end{document}